\documentclass{webofc}
\usepackage[varg]{txfonts}   % Web of Conferences font
\usepackage{xcolor}
\usepackage{soul}
\begin{document}
\title{3D Multi-system Bayesian Calibration with Energy Conservation to Study Rapidity-dependent Dynamics of Nuclear Collisions}

\author{\firstname{Andi} \lastname{Mankolli} 
\inst{1}\fnsep
\thanks{\email{andi.mankolli@vanderbilt.edu}} for the JETSCAPE Collaboration
        %\firstname{Second author} \lastname{Second author}\inst{2}\fnsep\thanks{\email{Mail address for second
             %author if necessary}} \and
        %\firstname{Third author} \lastname{Third author}\inst{3}\fnsep\thanks{\email{Mail address for last
             %author if necessary}}
        % etc.
}

\institute{Department of Physics and Astronomy, Vanderbilt
University, Nashville TN 37235 
%\and
           %the second here 
%\and
           %Last address
          }

\abstract{
  Considerable information about the early-stage dynamics of heavy-ion collisions is encoded in the rapidity dependence of measurements.
  To leverage the large amount of experimental data, we perform a systematic analysis using three-dimensional hydrodynamic simulations of multiple collision systems --- large and small, symmetric and asymmetric. 
  Specifically, we perform fully 3D multi-stage hydrodynamic simulations initialized by a parameterized model for rapidity-dependent energy deposition, which we calibrate on the hadron multiplicity and anisotropic flow coefficients. We utilize Bayesian inference to constrain properties of the early- and late-time dynamics of the system, and highlight the impact of enforcing global energy conservation in our 3D model. 
}
\maketitle
\section{Introduction}
\label{intro}

High-energy nucleus collisions at the Relativistic Heavy Ion Collider (RHIC) and the Large Hadron Collider (LHC) produce quark-gluon plasma (QGP).
As it cools, the QGP undergoes rapid 3-dimensional hydrodynamic expansion.
A popular and partly successful theoretical description of QGP evolution in heavy-ion collisions has been the longitudinal boost-invariant fluid, an approximate symmetry present in a significant region around mid-rapidity. 
Multi-stage models utilizing longitudinal boost invariance made use of 2D models of energy deposition in the initial state to compare with mid-rapidity measurements~\cite{Bernhard:2019bmu,PhysRevLett.126.242301,PhysRevLett.126.202301}. 
In recent years 3D models of the initial state, coupled to three dimensional viscous hydrodynamics, have been developed and tested 
against experimental data extending beyond mid-rapidity~\cite{Shen:2017bsr,PhysRevC.105.064905,PhysRevC.96.044912}. In this study, we use such a 3D model in conjunction with experimental data of high-energy collisions from four RHIC experiments in a Bayesian framework to quantify constraints on the initial state and the properties of the QGP.

In particular, we utilize
a 3D model in a Bayesian inference framework to study the effect of the temperature dependence of the specific shear and bulk viscosity coefficients of the QGP on dynamics at forward and backward rapidities. We do this in a multi-system calibration, using data from Au-Au and d-Au collisions at the same energy (200 GeV). We highlight the impact of using an initial-state
model featuring energy-momentum conservation, which propagates to the hydrodynamics
the incoming energy in the collision.
The present study serves as an exploratory first step toward a Bayesian system scan of all high-energy data probing the 3D dynamics of heavy-ion collisions at RHIC and eventually also the LHC.

\subsection{The model}
\label{sec-1}
A variety of 3D initial state models have been developed in recent years, including extensions of the 2D T$_R$ENTo and Monte Carlo Glauber models to three dimensions~\cite{PhysRevC.105.064905, Soeder:2023vdn}. In this study we use the 3D Glauber Model from Ref.~\cite{Shen:2017bsr,PhysRevC.105.064905}.
The positions of collisions in the transverse plane are determined using the same prescription as in the 2D Glauber model. At these positions, strings then extend in the longitudinal direction and are decelerated by a classical string tension~\cite{Shen:2017bsr,Shen:2017fnn}. The energy of the string, which is then transferred to the medium, is proportional to the energy lost during the collision. A rapidity loss is defined as a function of the incoming rapidity of the participants and piece-wise parameterized at three incoming rapidity points. A larger rapidity loss corresponds to a greater amount of energy deposited in the mid-rapidity region. The initial state model is used to initialize the 3D hydrodynamic evolution in MUSIC, after which the UrQMD hadronic afterburner simulates decays and scatterings. In past analyses using different initial condition models, a normalization parameter was used to scale the deposited energy density to fit final charged hadron multiplicities for different energies or collision systems. In this analysis, for a given event, our model is constrained to a precise energy density profile in 3D due to the conservation of energy, and therefore, such a scaling parameter is not applicable. In this way, the same set of parameters is varied for both d-Au and Au-Au. 

\section{Bayesian Analysis and Model Emulation}
\label{sec-2}

A Bayesian analysis enables us to obtain probability distributions for model parameters that are constrained by a particular set of experimental measurements and their corresponding uncertainties--that is, obtain posterior probability distributions from mildly informed priors. Our analysis includes 20 model parameters.
%and a correspondingly large number of MCMC walkers traversing the parameter space in thousands of steps. 
In order to make model predictions at such a large number of parameter sets, we make use of computationally-cheaper Gaussian Process Emulators (GPs) at the expense of introducing emulator uncertainty. After dimensional reduction of the observable space, GP surrogates are trained to reproduce the model predictions for a set of 10 principal components. The emulators are systematically validated against a set of model predictions for parameter sets not used in emulator training. Additionally, the analysis as a whole is validated by means of closure tests, in which the model predictions at given parameter sets are used as pseudo-experimental data. We look for posterior distributions of particularly sensitive parameters to converge on the "truth" values of the parameter set used for closure.

The experimental data used in this first iteration of a 3D analysis are the
charged hadron multiplicity and elliptic flow measurements as a function of pseudorapidity for various centrality bins in both Au-Au and d-Au collisions from the PHOBOS, BRAHMS, STAR, and PHENIX experiments~\cite{PhysRevLett.88.202301, PhysRevC.72.014904, PhysRevC.83.024913,PhysRevC.96.064905}. The different data sets extend to varying pseudorapidiy, though all go well beyond the mid-rapidity region. We use the 20-70\% and 0-5\% centrality bins for the $v_2$ in Au-Au and d-Au, respectively. Experimental uncertainties are taken to be uncorrelated across centrality and pseudorapidity bins.

\begin{table}[ht]
\centering
\begin{tabular}[t]{ccccc}
% Table content
\hline
\hline
\vspace{5pt}
System & \multicolumn{2}{c}{AuAu} & \multicolumn{2}{c}{dAu} \\
\hline
\vspace{5pt}
Observable & $v_2$($\eta$) & $dN_{ch}$/d$\eta$ & $v_2$($\eta$) & $dN_{ch}$/d$\eta$\\
%\hline
\hline
No. of data points & 21 & 120 & 24 & 162 \\
\hline \hline
\end{tabular}
\caption{System and observable breakdown of the experimental data used in this analysis.}
\vspace{-10pt}
\end{table}

\section{Constraints on Initial State and Transport Properties}
\label{sec-3}
\vspace{-5pt}

\begin{figure}[h]
\centering
\includegraphics[width=6cm,clip]{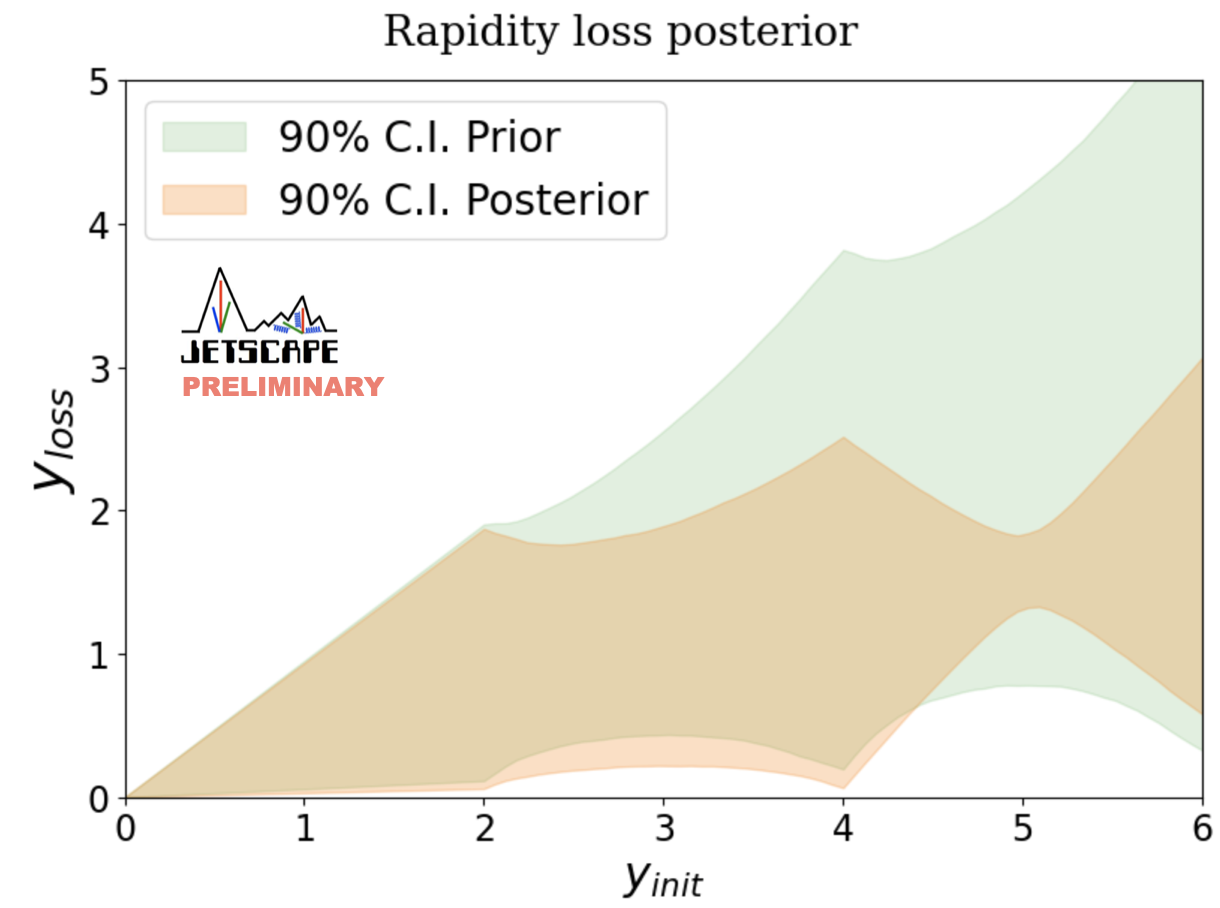}
\includegraphics[width=6cm,clip]{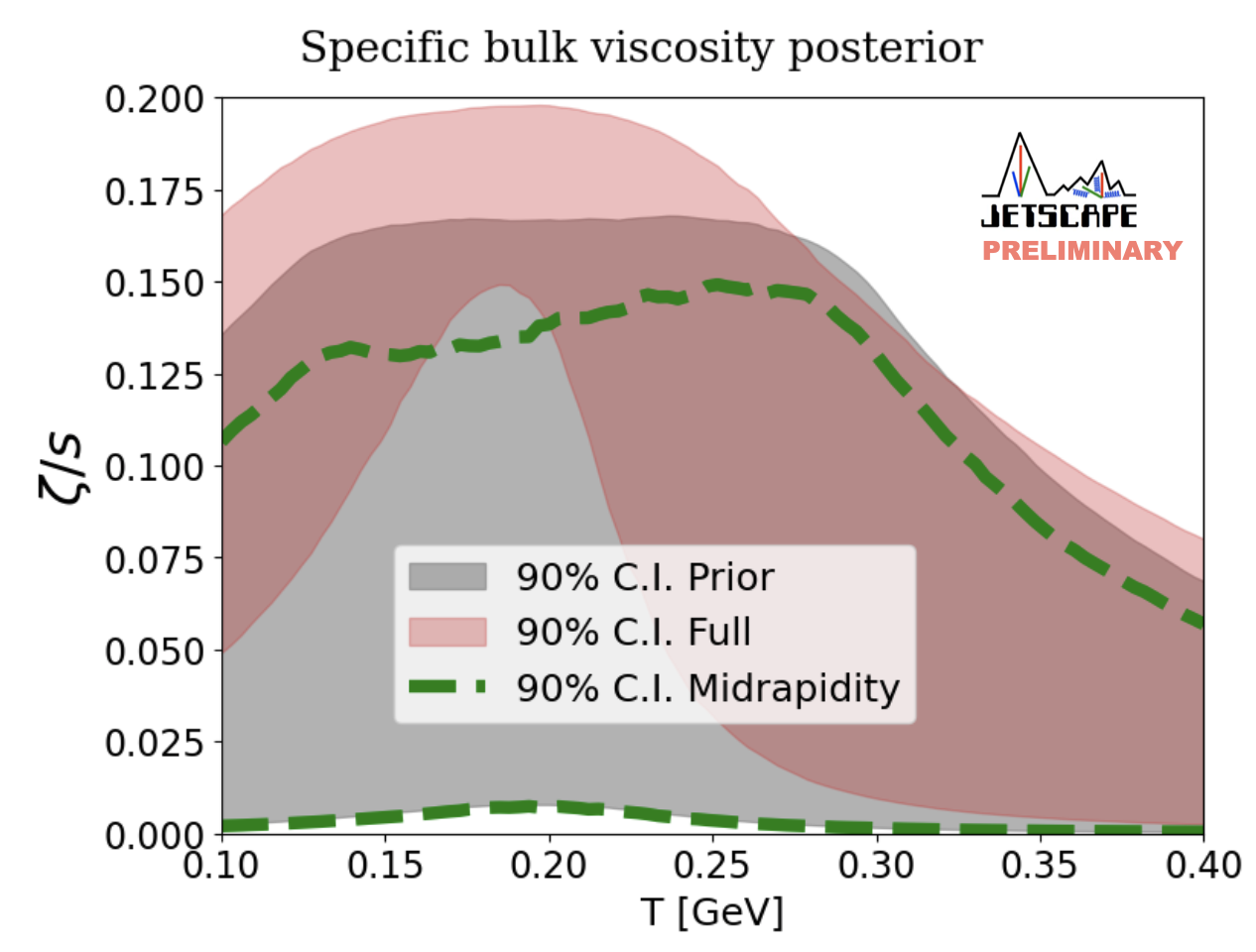}
\caption{The 90\% confidence intervals (C.I.) for the prior and posterior of the rapidity loss as a function of the incoming rapidity (left) and specific bulk viscosity as a function of temperature (right).}
\label{fig-1}      
\vspace{-10pt}
\end{figure}

Constraints on the rapidity loss parameters are shown on the left side of Fig.\ref{fig-1}. As expected, the strongest constraints are in the incoming rapidity region corresponding to the beam rapidity at 200 GeV. We verified that most observables are strongly sensitive to the value of rapidity loss parameters at 4 and 6 units, consistent with the recent Bayesian analysis in ~\cite{Shen:2023awv}. The analysis also shows little constraining power in regions of lower incoming beam rapidity.

The data also indicate a preference for a specified large bulk viscosity, as shown on the right panel of Fig.\ref{fig-1}.
The preferred viscosity is so large that it extends beyond the 90\% prior for $\zeta/s$.
We verified (not shown) that a large bulk viscosity is found whether one uses on only Au-Au or only d-Au data.

For completeness, we repeated the Bayesian analysis using only mid-rapidity measurements ($|\eta|<1$). The result is shown as the green dashed line in the right panel of Fig.\ref{fig-1}. We find that one can describe mid-rapidity data with small values of $\zeta/s$, indicating that the large values of the specific bulk viscosity found in the 3D analysis are truly from the rapidity-dependent measurements. In fact, the constraints on the bulk viscosity found with the mid-rapidity measurements are generally consistent with the ones found in a previous JETSCAPE calibration~\cite{PhysRevLett.126.242301}.
This is not obvious, since the model and data sets used in the present work and in Ref~\cite{PhysRevLett.126.242301} are rather different; further verification will be necessary to ascertain that this conclusion remains the same after more measurements are added to the analysis.

% figure~\ref{fig-1}

\subsection{Normalization of Initial State Energy Deposition}
\label{sec-4}

To successfully describe charged hadron multiplicity measurements,
previous analyses scaled the initial state energy density by a normalization parameter that may be fit to each system or energy independently. In the absence of such an explicit normalization in a Bayesian analysis, the remaining parameters sensitive to the multiplicity compensate for a good fit to the multiplicity data in different systems. The most sensitive are the rapidity loss parameters, which define a 3D energy profile. Another one is the bulk viscosity, which affects the final multiplicities through viscous heating of the system during the hydrodynamic evolution. When the model is constrained only at mid-rapidity, the rapidity loss acts in effect as a normalization parameter to match with the multiplicity at mid-rapidity. Indeed, as we noted, when we elect to use only mid-rapidity data in our analysis, we obtain constraints on the bulk viscosity similar to those of previous 2D analyses. When forward/backward rapidity data is included in the analysis, however, the rapidity loss is constrained by global conservation of energy, and an absolute scale degree of freedom is no longer allowed. To optimize the absolute scale to match the magnitude of the multiplicity, the bulk viscosity at low temperature needs to be large. The results we see in this work are consistent with this interpretation. By contrast, we did not find strong constraints on the shear viscosity (not shown) even when rapidity-dependent observables are included.

\section{Outlook}
\label{sec-5}

In this work, we have performed a multi-system Bayesian comparison of rapidity-dependent measurements of charged hadron multiplicities and elliptic flow with a 3D model of the initial state and hydrodynamic evolution of high-energy nuclear collisions. We have noted the strong constraints on the rapidity loss and the system dependence of the calibration. We have, in particular, seen indications of a preference for a large bulk viscosity and discussed the role of energy-momentum conservation in a 3D model on this result. The JETSCAPE collaboration is actively working on 3D calibrations using a more extensive set of available experimental data.   

In addition, a 3D calibration of the properties of the QGP fluid throughout its space-time evolution utilizing all available soft hadron data is a crucial baseline for further studies that couple hard physics to a hydrodynamic simulation for a complete description of heavy ion collisions in realistic jet+hydro simulations, a long-standing goal of the community. 

\vspace{8pt}
\textit{Acknowledgments.} This work was supported in part by the National Science Foundation (NSF) within the framework of the JETSCAPE collaboration, under grant number OAC-2004571 (CSSI:X-SCAPE) through a subcontract with Wayne State University and used allocations on the Anvil and Stampede2 supercomputer clusters at RCAC and TACC.
\vspace{-2pt}
%
% BibTeX or Biber users please use (the style is already called in the class, ensure that the "woc.bst" style is in your local directory)
\bibliography{template.bib}
%\begin{thebibliography}{}
%\bibitem{RefJ1}
%Journal Author, Journal \textbf{Volume}, page numbers (year)
%Journal Author, Journal \textbf{Volume}, page numbers (year)
%Journal Author, Journal \textbf{Volume}, page numbers (year)

%\end{thebibliography}

\end{document}